# Reputation dependent pricing strategy – analysis based on a Chinese C2C marketplace

**Zehao Chen [1], Yanchen Zhu[2], Tianyang Shen[3], Yufan Ye[4]**

*1 International Business School Suzhou, Xi-an Jiaotong Liverpool University, Suzhou, Jiangsu, 215123, China*
*2 School of Chemistry, The University of Edinburgh, Edinburgh, Scotland, EH89YL, Britain*
*3 Mathematical Institute, University of Oxford, Oxford, OX2 6GG, Britain*
*4 Foreign Languages School, Zhejiang University, Hangzhou, Zhejiang, 310012, China*

*[*]Corresponding author. Email: Zehao.Chen18@student.xjtlu.edu.cn*

***ABSTRACT.***

*Most online markets establish reputation systems to assist building trust between sellers and buyers. Sellers' reputations not only provide guidelines for buyers but may also inform sellers their optimal pricing strategy. In this research, we assumed two types of buyers – informed buyers and uninformed buyers. Informed buyers know more about the reputation about the seller but may incur a search cost. Then we developed a benchmark model and a competition model. We found that high reputation sellers and low reputation sellers adapt different pricing strategy depending on the informativeness of buyers and the competition among sellers. With a large proportion of well-informed buyers, high reputation sellers may charge lower price than low reputation sellers, i.e., a negative price premium effect, in contrast to conclusions of some previous studies. Empirical findings were in consistence with our theoretical models. We collected data of five categories of products, televisions, laptops, cosmetics, shoes, and beverages, from Taobao – a leading C2C Chinese e-market. Negative price premium effect was observed for TVs, laptops, and cosmetics; price premium effect was observed for beverages; no significant trend was observed for shoes. We infer product value and market complexity are the main factors of buyer informativeness.*

## 1. Introduction

With the development of the internet, the emergence of big data, blockchain and other technologies has had a huge impact on people's lives. Until January 2021, there were 5.22 billion internet users around the world. By December 2020, the number of Chinese internet users had reached 989 million. Online market plays an important role in the 21st century because of the increasing number of busy people, loaded with hectic schedule. In this context online shopping became a convenient and popular shopping mode. [1] With the rapid development of internet economy and the reduction of operating costs, the online market has attracted a large number of merchants, and the modern e-commerce market has gradually matured. However, at present, it is because of the entry of a large number of e-commerce enterprises into the network market that the competition among similar e-commerce enterprises becomes increasingly fierce and the homogenization phenomenon becomes more serious. The main means of competition, such as price reduction and malicious bad comments, are used to confuse consumers and disrupt the market order. At the same time, there are more and more problems have appeared while the growing of e-commerce, like reputation inflation and internet fraud. [2] This paper focus on the reputation system in the e-commerce market for in-depth research.
Reputation systems are essential to e-commerce markets and online platforms. [3] Reputation plays a crucial role in the success of e-commerce. It is necessary to evaluate transactions in a timely and a robust manner where sellers and buyers do not meet in person, so only previous feedbacks can reflect seller’s quality.[4] Reputation systems serve as core building blocks in different platforms such as eBay and Taobao.[5] E-commerce, as a new way of commercial trade, makes use of the Internet and advanced digital media techniques to carry out various commercial trade, and establishes a new economic order. The premise of mutual exchange is to recognize each other's reputation. Most online marketplace transactions are regulated by the reputation system, which allows consumers to rate goods and services by scores and comments. [6] The basis of this credit rating is the choice of contractual norms in the pursuit of long-term benefits in long-term economic transactions. Reputation, as a product of commercial transactions, is of great importance to promoting the development of market economy. Online word of mouth has a positive impact on purchase intention, while trust and price fairness play a mediating role between reputation and purchase intention. [7] The improvement of the reputation system can promote the ecological development of the e-commerce market, maintain the market order, and provide a good market environment for sellers and buyers. There are many factors influencing ratings, and there have been records that people use text sentiment analysis technology to deeply dig the text information of e-commerce reviews to reflect the attributes and attitudes of consumers. The usefulness of a single scoring mechanism will be limited to a certain extent by text evaluation. [8] As price is visual and suitable for quantitative analysis, we focus on the relationship between prices and reputations.

In previous literature, S.L. Ba and P.A. Pavlou argued that buyers are likely to pay a premium to reputable sellers. Therefore, reputable sellers should charge a relatively high price. However, research of S.L. Ba, J. Stallaert and Z. J. Zhang showed opposed results. [9] For example, if a seller who pay more attention to price performance or do not understand the quality of products from different suppliers. [10] Meanwhile, K. Baylis, J.M. Perloff point out that sometimes "good" online retailers charge relatively low prices and provide excellent service, while "bad" online retailers charge relatively high prices and provide poor service. [11] We found that there were certain contradictions between the conclusions of many previous literatures.

The rest of the paper is organized as follows. The second part summarizes the relevant literature. The third part develops the main theoretical models. An empirical analysis is presented in part four.

## 2. Literature Review

### 2.1. Online market prices are decentralized

Because of advances in Internet, the number of sellers entering the e-commerce market is increasing, and the competition in the online market is fierce. Big data analysis is especially widely used in the electronic market and has a profound impact on the development of global e-commerce. [12] It's generally believed that online market can cut the costs of searching, make the transmission of information efficiency and transparency. Electronic markets will eventually make trading friction become less, that is, online transactions will promote the development of e-commerce, the same commodity market prices tend to be the same, and the market price decreases the discrete situation between different vendors or even disappear. [13] According to the Bertrand model, in an ideal frictionless market, buyers can buy goods from sellers at the lowest price. [14] According to the law of one price, all sellers shall set the same price. [15] However, in contrast to the theoretical predictions, Li Weiwei and other researchers confirmed in network sales did not lead to a trade friction, on the basis of consumer demand from the perspective of consumer purchase decision behavior analysis, found that website content, the enterprise brand image, transaction security and privacy security, distribution and after-sales service and so on non-price factors influence the purchase decision-making behavior, Furthermore, consumers' demands affect sellers' price strategies, which makes friction less transaction difficult to achieve. [13], that is, the price in the online market is dispersed.

### 2.2. Factors Affecting price dispersion in online markets

#### 2.2.1. Reputation of the Seller

In e-commerce platforms, trust has a significant impact on transactions. A seller's reputation is usually calculated by buyer reviews and reviews (e.g., five-star reviews and written reviews on Taobao). There are many factors influencing the evaluation, such as service and logistics. affecting the rate of reputation. [18] Higher product sales and seller reputation scores may also indicate that the seller has higher reliability, accurate product descriptions, and better logistics and after-sales service. [16] Therefore, in general, buyers tend to pay higher prices to sellers with low risk, good service and good reputation to protect their own interests [17], thus affecting price dispersion in the e-commerce market.

#### 2.2.2. Buyer's Knowledge

Although search costs in the online market have been greatly reduced due to the rapid development of the Internet, buyers may still have to afford search costs in the online market. [19] They have different online shopping experience, different searching skills and different level of wealth, so the search may be varied.[20] Initially we simply think that the existence of unsuspecting buyers' market information for nothing, but in fact there is no credit to companies do not know the buyer. Hence, according to the different search costs, buyers can be divided into informed buyers and uninformed buyers. Informed buyers are those who can search and compare different products to get the information of sellers and buy with the highest utility while uninformed buyers may purchase the product immediately as long as the products offer positive effects [21]. Uninformed buyer will conduct a limited search[22]. Buyers' different cognition of the same product may be used by merchants to carry out different pricing strategies, thus affecting the price distribution in the electronic market.

#### 2.3. Information Asymmetry between Buyers and Sellers

In the current market, data and information has become a hot source of competition among sellers. [23] In contemporary society, the low-cost advantage of e-commerce market provides great convenience for the seller of inferior goods to enter. In addition, under the advantage of low cost, sellers may also take the advantage of information asymmetry to provide buyers with untrue product quality information. Although a large amount of information is conducive to buyers' decision-making, but at the same time, the inaccurate information will also lead to buyers to make wrong judgments, and the information asymmetry in the e-commerce market will intensify. [24] This just confirms the market effect of lemons proposed by Akerlof in 1970, which refers to the phenomenon that inferior goods drive out normal good due to information asymmetry, which leads to market disorder. [25] On e-commerce platforms, the impact of the lemon market is likely to lead to higher prices for low-reputation sellers.

#### 2.4. The Relationship between Reputation and Price in E-commerce

It has been pointed out that in the e-commerce market, reputation and price show a positive correlation, that is, a higher feedback score has a positive impact on price. [21] However, there are also literatures showing that reputation and price are inversely proportional. Our study provides an explanation for the contradictory findings in the previous literature.

## 3. Model

#### 3.1. Benchmark model: market with one low-reputation seller and one high-reputation seller

Assume a market only contains two sellers selling one identical product to n (n∈ N) buyers.[26] The two sellers were described by i, i∈{L, H}. $r_i$ stands for the reputation of the two sellers and 0<$r_L$<$r_H$<1. Therefore, $1-r_L$ and $1-r_H$ stands for the risk when making deal with the low reputation seller and high reputation seller. The product cost is c for sellers and the product price pi is charged by sellers. Sellers only choose to sell the product when they can get non-negative profit(pi-c≥0). Buyers achieve the utility u for the product. Consider the search cost for the informed buyers is k and uninformed buyers do not have search cost. Hence, $\frac{k}{u}$ ($0<\frac{k}{u}<1$) represents the proportion of the uninformed buyers and $1-\frac{k}{u}$ represents the proportion of the informed buyers. Based on the Liu's model [22], in the transaction, the expected utility of consumer is $r_i u - p_i$.

Similar to Salop and Stiglitz [27], we suppose that an informed buyer will do the search and compare the products from different sellers and buy the product with the highest expected utility given by specific seller. For uninformed buyers, they choose one seller to purchase the product at random. The buyer will only make the transaction with sellers when they achieve non-negative expected utility. Through calculation, for low-reputation seller L, the price domain is $p_L \in [c + \frac{k}{2u-k}(r_L u - c), r_L u]$. For high rating seller H, the price domain is $p_H \in [c + \frac{k}{2u-k}(r_L u - c) + (r_H - r_L)u, r_H u]1$ We define $p_H = c + \frac{k}{2u-k}(r_L u - c) + (r_H - r_L)u$, $\widehat{p_H} = r_H u$, $p_L = c + \frac{k}{2u-k}(r_L u - c)$, $\widehat{p_L} = r_L u$.

In this model, $\frac{k}{u}n$ uninformed buyers are equally shared by the two sellers. The two sellers will compete for the $(1 - \frac{k}{u})n$ buyers. The buyers will be attracted by the seller i who gives them the higher utility. Hence, the seller i will win the whole market of informed buyers with a profit of $(p_i - c)(\frac{u-k}{u}n + \frac{(2u-k)n}{2u}) = (p_i - c)(u - \frac{kn}{2u})$. In terms of the other seller (-i), he will only sell their product to the half of the uninformed buyers and its profit is $\frac{kn}{2u}(p_{-i} - c)$. If the two sellers give the same expected utility to the buyers, they will equally share the informed buyers with the profit of $\frac{n}{2}(p_i - c)$. Hence, in order to maximize his profit, seller i may find the optimal price:

$$\mathrm{Max}_{p_i} \left(u - \frac{kn}{2u}\right)(p_i - c)P\left(r_i u - p_i > r_{(-i)}u - p_{(-i)}\right) + \frac{n}{2}(p_i - c)P\left(r_i u - p_i = r_{(-i)}u - p_{(-i)}\right) + \left(\frac{kn}{2u}\right)(p_i - c)P\left(r_i u - p_i < r_{(-i)}u - p_{(-i)}\right)$$

Therefore, similar to Liu [22], pure strategy equilibrium may not be existed in this model. To explain, the uninformed buyers give the motivation for sellers to set a relative high price but the informed buyers may make the seller set a low price. Due to the co-existence of the uninformed buyers and informed buyers, the seller cannot set a pure price to satisfy both informed buyers' and uninformed buyers' utility. Hence, sellers randomize their prices at the end. Let $\overline{F_i}^*(p_i^*)$ represents the probability density function (PDF) and $F_i^*(p_i^*)$ represents the cumulative distribution function (CDF) of equilibrium prices for seller i. For each of the two sellers, regarding the other seller's mixed pricing strategy, the equilibrium price will exist when the profit is the same regardless of the price set in the given range. The logic of mixed pricing strategy will be explained in Proposition 1.

**Proposition1:**

$\{p_L^*, p_H^*\}$ is the equilibrium price following a mixed strategy:

$p_L^* \in [p_L, \widehat{p_L}]$, $\overline{F_L}^*(p_L^*) = \frac{(2u-k)(r_H u - c) - 2(u-k)(r_L u - c)}{2(u-k)(p_L^* + (r_H - r_L)u - c)^2}$ and $P(p_L = \widehat{p_L}) = \frac{(r_H - r_L)u}{r_H u - c}$

$p_H^* \in [p_H, \widehat{p_H})$, and $\overline{F_H}^*(p_H^*) = \frac{k(r_L u - c)}{2(u-k)(p_H^* - (r_H - r_L)u - c)^2}$

We can discover that for the low rating seller, he may set the price $\widehat{p_L}$ with the possibility $P(p_L = \widehat{p_L})$. It is reasonable because the low reputation seller cannot win the high reputation seller in getting the market of informed buyers. Hence, the low reputation seller gives up all the informed buyers. Instead, the low reputation seller sells the product to the uninformed buyers and set the product price as high as possible to maximize his profit.

Figure 1: Equilibrium CDF of benchmark model

In the following examples, u=2, c=1, $r_H$=0.9 and $r_L$=0.8, the uninformed buyer's proportion is $\frac{k}{u}$ =0.7 in figure 1a) and 0.3 in figure 1b)

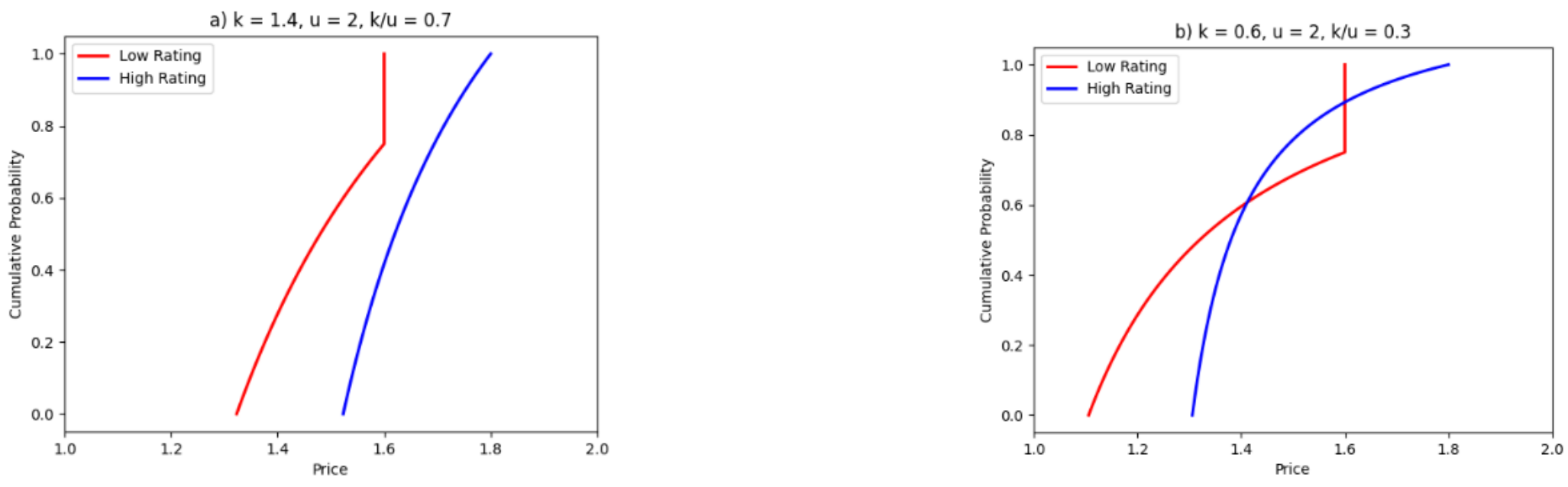

Figure 1 gives two numerical examples based on the relative domain of the pricing strategy of the two sellers. In figure 1a), the CDF of the low reputation seller is above the CDF of the high reputation seller. $F_H^*(p_H^*) \leq F_L^*(p_L^*)$. This means there exist a price premium in the expected price between the high reputation seller and the low reputation seller. In figure 1b), $(1-\frac{k}{u})$ increases, which means the proportion of high-reputation seller increases. This situation makes the market more competitive. Hence, both sellers reduce price at lower bound $(\frac{d_{p_i}}{d_k} > 0)$. The probability of charging a low price also increases $(\frac{d_{F^*_i}}{d_k} > 0)$. Therefore, the expected prices of both sellers decrease $(\frac{d_{E(p_i)}}{d_k} \geq 0)$. Furthermore, with the increasing proportion of informed buyers, the price premium of the high-reputation seller disappear. Since the expected price of low reputation seller drops much slowly than that of high-reputation seller $(\frac{d(E(p^*_H)-E(p^*_L))}{d_k} > 0)$, as the profit of low-reputation seller depends on the uninformed buyers, not the informed buyers.

3.2 Competition model: market with two-low reputation sellers and two-high reputation sellers

Assume $H_1$ and $H_2$ stand for high reputation sellers and they have the same reputation of $r_H$. $L_1$ and $L_2$ stand for low reputation sellers and they have the same reputation of $r_L$. $p_{ij}$ represents the price set by seller ij, i∈{L,H} and j ∈{1,2}. Each of the high reputation seller not only competes with the sellers with low reputation but also compete with the other high reputation seller. Each of the low reputation seller not only compete with the sellers of high reputation but also compete with the other low reputation seller. Through calculation, the price domain of seller ij is pij∈[c +

$\frac{k}{4u-3k}(r_i u - c), r_i u]$. We define pij = $c + \frac{k}{4u-3k}(r_i u - c)$ and $\widehat{p_{ij}} = r_i u$

In this model, we assume the buyers are rational, which means they may buy the product of the highest utility. Each seller may have two strategies. The first strategy is that among the four sellers, if one seller set the lowest price occasionally, he can sell the product to every informed buyers and a quarter of uninformed buyers. The second strategy is that the seller can give up the market of informed buyers and charge the highest possible price to the targeted quarter of uninformed buyers. In the benchmark model, both sellers can sell their product at the lower-bound price and both of them give consumers the same expected utility. However, in this competition model, there is no room for the low reputation seller to decrease the lower-bound price. Hence, giving up the first strategy, the low reputation seller can only choose the second strategy. Let $\overline{F_{ij}}^*(p_{ij}^*)$ represents the probability density function (PDF) and $F_{ij}^*(p_{ij}^*)$ represents the cumulative distribution function (CDF) of equilibrium prices for seller ij. The logic of mixed pricing strategy will be explained in Proposition2.

**Proposition 2:**

$\{p_{Lj}^*, p_{Hj}^*\}$ is the equilibrium price following a mixed strategy:

$p^*_{Hj} \in [p_{Hj}, \widehat{p_{Hj}}]$ and $\overline{F_{Hj}}^*(p_{Hj}^*) = \frac{k(r_H u - c)}{4(u-k)(p_{Hj}^* - c)}$

$p^*_{Lj} = \widehat{p_{Lj}}$, j=1, 2

Figure 2 shows two numerical instances. [3] From the graphs, it can be easily observed that whatever price the high reputation seller set, low reputation seller may not change his price.

Figure 2: Equilibrium CDF in the competition model

In the following two examples, u=2, c=1, $r_H$=0.9 and $r_L$=0.8, the uninformed buyer's proportion is $\frac{k}{u} = 0.7$ in figure 2a) and 0.3 in figure 2b)

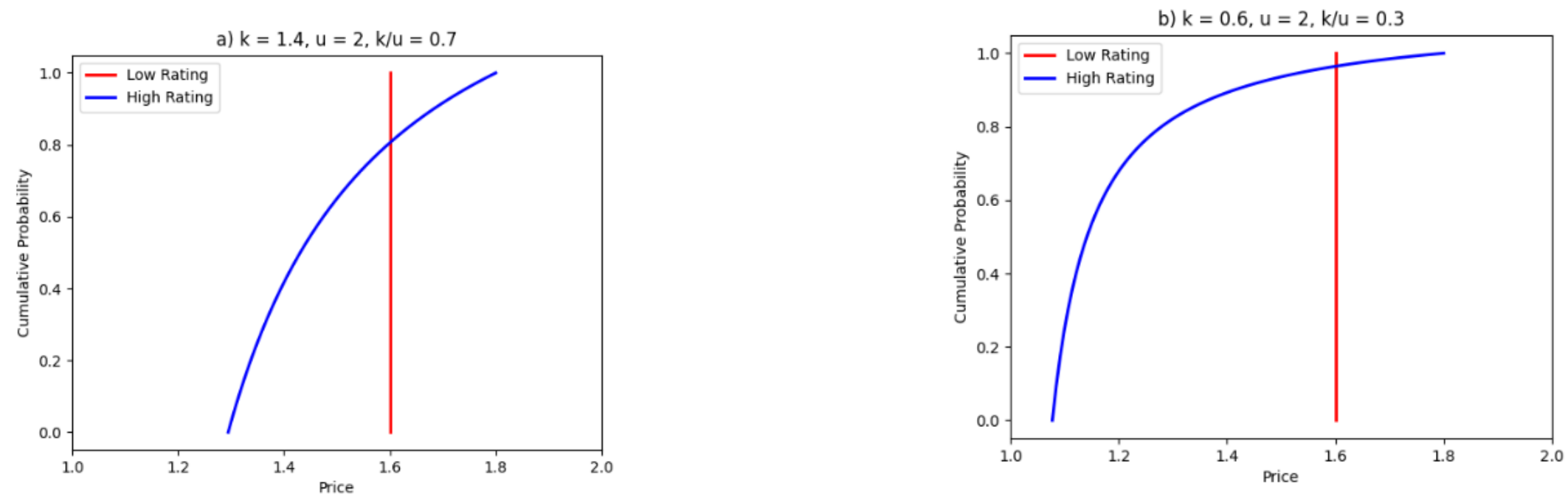

When more sellers exist in the market, the pricing strategy of low-reputation sellers do not change. They always choose the highest possible price to sell the product to uninformed buyers.

In terms of the high reputation seller, their lower bound price may be reduced ($\frac{d_{p_{Hj}}}{d_k} > 0$). The probability of charging a low price also increases ($\frac{d_{F^*Hj}}{d_k} > 0$). The expected prices decrease

$$\frac{dE({p^*}_{Hj})}{d_k} \geq 0$$

### 3.3 Price comparison

In both reputation model the benchmark model, the pricing strategy of the low reputation seller may not be changed. Low reputation seller sometimes may charge higher product price than high reputation seller. Furthermore, we need to understand two questions. (1) Whose expected price may be higher, the high reputation seller's or the low reputation seller's? (2) When the competition between the sellers intensifies, will the seller set a lower price?

#### 3.3.1 Price relationship between low-reputation sellers and high-reputation sellers

Inference 1.

In the benchmark model, the price of high rating seller and the price of low rating seller may be equal at $k_1^*$. When $k > k_1^*$, the high rating seller's expected price is higher. When $k < k_1^*$the low rating seller's price is higher.

Inference 2.

In the competition model, the price of high rating seller and the price of low rating seller may be equal at $k_2^*$. When $k > k_2^*$, the high rating seller's expected price is higher. When $k < k_2^*$ the low rating seller's price is higher.

Both inferences illustrate that if the proportion of the high rating sellers is high enough, the expected price of the low rating sellers may exceed high rating seller's expected price. Figure3 and Figure4 give numerical examples to inference1 and inference2 respectively. Comparing figure3 and figure4, we can find that the probability of the negative price premium to appear is higher in the competition model when $k_1^* < k_2^*$. Later appendix will give a formal proof.

Figure 3:

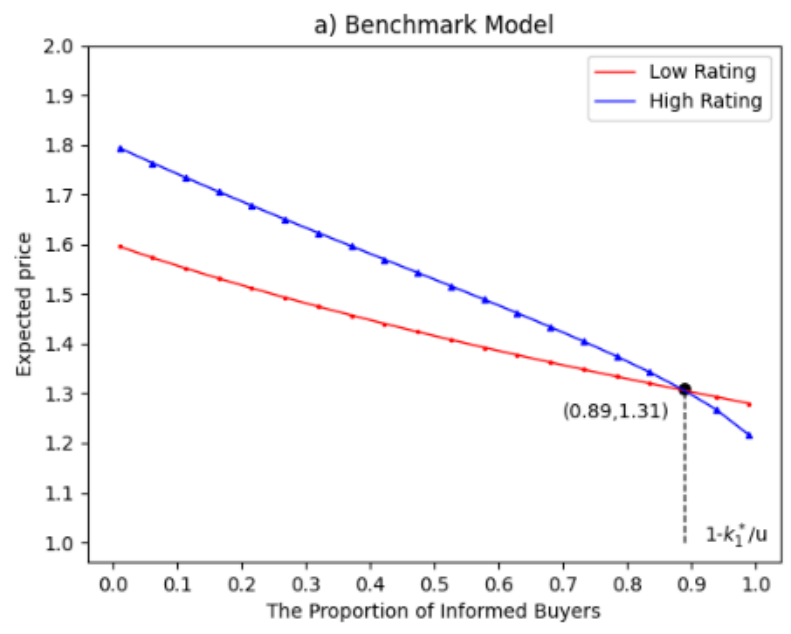

Figure 4:

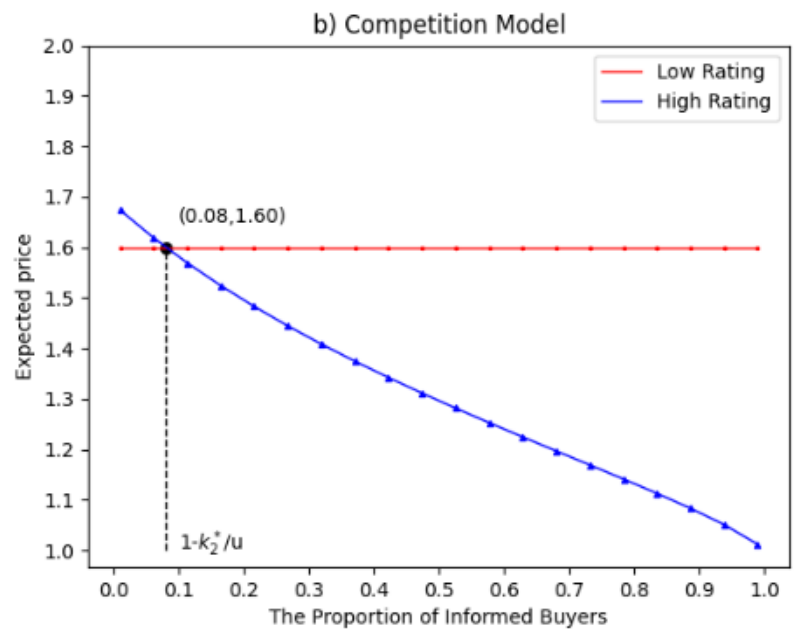

#### 3.3.2 Competition effect

Since $F_H{}^*({p_H}^*) \leq F_{Hj}{}^*({p_{Hj}}^*)$, the expected price of high rating seller in the benchmark model is higher than in the competition model because one more seller enter the market and the two high rating seller compete with each other. In terms of the competition model, the price of low rating seller is $r_L u$, which is equal to the highest possible value of the price domain of low rating seller in the benchmark model. However, in the model, the low rating seller cannot set the product price at upper bound of the domain. To conclude, the existence of competition reduces the price of high rating seller and increases the price of low rating seller price a bit. In addition, it is easy to prove $k_1^* < k_2^*$.

Inference 3:
When the market becomes more competitive, the probability that the high reputation seller sets lower price will increase.

In conclusion, when the market exist competition, sellers of different reputation level may have different pricing strategies: when more and more sellers enter into the online market, in order to get the market of the informed buyers, the high reputation sellers may reduce the product price. In terms of the low reputation sellers, they may choose the highest possible price for the uninformed buyers and give up the market of all informed buyers.

## 4. Empirical analysis

### 4.1 Data collection

There are several popular Chinese online markets such as JingDong and SuNing in addition to TaoBao. We choose TaoBao as our source of data as TaoBao is a leading Chinese C2C online marketplace with a diversity of products and sellers. For example, when searching for a lipstick, TaoBao shows 2784 results, compared to 1416 on JingDong and none on SuNing. TaoBao's latest reputation system was initiated in 2009 which consists of binary rating system and component rating system [28]. The binary rating system allows buyers to rate a product through positive, neutral, or negative evaluation with text description. Simple aggregation is applied to each of the ratings and the result is shown in the product page. The component rating system establishes three detailed indicators, i.e., description accuracy, customer service, and shipping, to rate a seller depending on customers' level of satisfaction. The average of ratings within 6 months is referred to as dynamic shop rating and is shown on all the product pages from this shop. One buyer was able to rate one shop no more than three times a month after his purchase, which decreased the likelihood of 'score farming'. As the reputation of a seller is the average satisfaction level he gets, dynamic shop rating reflects seller reputation in Taobao. In this research, the rating of each seller is defined by the average of dynamic shop ratings.
Our data were obtained from TaoBao in May 2021. A web crawler was developed to automatically grab all the offerings of 45 products in 5 categories from various sellers. The web crawler 'input' the keyword of each product and 'clicked' into offering pages from search results. Then following information were collected: the dynamic shop ratings, cumulative comments, name of seller, location of seller, title of offering, and 'TaoBao price'. 'TaoBao price' was the unconditional promotion price set by seller. In most cases, it was lower than the 'original price', which was also presented. As 'TaoBao price' could be a selling strategy to attract potential customers and it was the real transaction price, only 'TaoBao price' was considered and referred to as 'price' hereinafter.

As the raw data may contain unwanted information, e.g., different products with similar name, raw data were cleaned to guarantee the offerings of each product were homogeneous by following steps:

(i) We dropped sellers with no rating.

(ii) We removed used or refurbished offerings.

(iii) We removed offerings from official sellers as off-line reputation may contribute to their selling strategy.

(iv) We removed extreme offering prices which are 5 times standard deviations above or below the average price of the same product. For example, an offering of TV charged 100 yuan, which was obviously unreasonable compared to the average price. These data might be listed there by error.

(v) We refined the results by searching for keywords of one specific configuration because multiple prices for different configurations might be presented on one offering page. For instance, in terms of 'iPhone 12 pro max', products with 128G, 256G, and 512G storage might occur on the same page. After refinement, offerings with 256G storage were recorded while others were dropped.

(vi) If there were multiple choices of colors on one offering page, e.g., iPhone 12 pro max in graphite and silver, their prices were averaged and rounded to an integer.

**Table 1:** Sample size and price dispersion of cleaned data

| Category | Sample size | | Offering price | | Standardized price | |
|---|---|---|---|---|---|---|
| | Products | Offerings | Max | Min | Max | Min |
| TVs | 9 | 329 | 13599 | 1849 | 4.63 | -4.38 |
| Laptops | 14 | 919 | 15719 | 2549 | 4.20 | -3.87 |
| Cosmetics | 6 | 834 | 341 | 162 | 3.65 | -2.72 |
| Shoes | 9 | 945 | 2405 | 264 | 3.87 | -3.24 |
| Beverages | 7 | 773 | 87 | 18 | 4.93 | -2.26 |

**Table 2:** Rating dispersion of cleaned data

| Category | Rating | | | | | The number of rating | |
|---|---|---|---|---|---|---|---|
| | Ave | Median | SD | Max | Min | >4.50 | >4.00 |
| TVs | 4.700 | 4.90 | 0.753 | 5.00 | 1.00 | 303(92.1%) | 309(93.9%) |
| Laptops | 4.816 | 4.80 | 0.063 | 4.93 | 4.70 | 907(98.7%) | 917(99.8%) |
| Cosmetics | 4.871 | 4.90 | 0.074 | 5.00 | 4.43 | 595(71.3%) | 596(71.4%) |
| Shoes | 4.788 | 4.80 | 0.156 | 5.00 | 3.33 | 922(97.6%) | 932(98.6%) |
| Beverages | 4.789 | 4.80 | 0.121 | 5.00 | 3.93 | 734(95.0%) | 761(98.4%) |

We chose 45 products from categories of television, laptops, cosmetics, shoes, and beverages. There are numerous dealers in addition to official sellers offering products of these categories, which results in the diversity of pricing and rating of sellers. **Table 1 and Table 2** were the statistical summary. There were 3700 offerings after the refinement. But note that ratings were predominantly concentrated in high score region. 95.0% of the ratings were above 4.0 and 93.5% of them were above 4.5.

## 4.2 Negative price premium effect

We used a multiple linear regression model to test the potential relationship between sales volume, number of comments, rating, and price, based on following equation:

$$y_i = \beta_0 + \beta_1 S_i + \beta_2 C_i + \beta_3 R_i + \varepsilon_i$$

Where yi is the standardized price; S is the sales volume; C is the number of comments; R is the rating. $\beta_0$ is a constant. $\varepsilon_i$ is the error term. These factors are taken into consideration because they are assumed to influence the pricing strategy of seller as they are quantitative indicators that potential buyers know when browsing for offerings. Estimated parameters and significance are listed in **Table 4**.

**Table 4:** Results of multiple linear regression

| | $\beta_0$ | $\beta_1$ | $\beta_2$ | $\beta_3$ |
|---|---|---|---|---|
| TVs | $0.77^{***}$ | $2.25 \times 10^{-3}$ | $-6.02 \times 10^{-4}$ | $-0.17^{**}$ |
| Laptops | $6.54^{***}$ | $4.03 \times 10^{-5}$ | $-6.41 \times 10^{-5}$ | $-1.32^{***}$ |
| Cosmetics | $19.62^{***}$ | $6.03 \times 10^{-5}$ | $7.46 \times 10^{-6}$ | $-4.08^{***}$ |
| Shoes | $-1.31^{*}$ | $3.07 \times 10^{-4}$ | $-6.70 \times 10^{-5}$ | $0.27^{*}$ |
| Beverages | $-14.71^{***}$ | $-4.87 \times 10^{-6}$ | $8.17 \times 10^{-7}$ | $3.07^{***}$ |

Note: ***, **, * denote that the null hypothesis of zero is rejected at the 1%, 5%, and 10% significance levels respectively.

The rating of seller is significantly correlated to price. Meanwhile, sales volume and comments number are irrelevant variables with small coefficients. Based on the knowledge that rating is the main factor of pricing strategy, we examined the pricing strategy of sellers with different rating levels.

Firstly, to make the price among different products in the same category comparable, offering prices were standardized. Then the median of seller's rating was calculated. All sellers were classified into two sub-categories. The sellers who possessed a rating less than median rating were denoted as 'low', where the sellers with a rating no less than median rating were denoted as 'high'. The average (AVE) and standard deviation (SD) of standardized price of sub-categories were listed in **Table 3**. The number of offerings with positive, negative, and zero standardized price was also listed. Finally, a two-tailed t-test was used to compare the average standardized prices of low- and high-rating sellers. The assumptions and results were listed in the last two columns.

**Table 3:** Statistics of sub-categories

| Category | Rating | Sample size | Standardized price | | The number of offerings | | | t-Test | |
|---|---|---|---|---|---|---|---|---|---|
| | | | AVE | SD | > 0 | < 0 | = 0 | Assumption | Sig. |
| TVs | High | 182 | -0.22 | 0.78 | 54 | 128 | 0 | $E(p^*_L) > E(p^*_H)$ | *** |
| | Low | 147 | 0.28 | 1.13 | 73 | 74 | 0 | | |
| Laptops | High | 523 | -0.16 | 0.90 | 219 | 313 | 0 | $E(p^*_L) > E(p^*_H)$ | *** |
| | Low | 389 | 0.22 | 1.06 | 225 | 162 | 0 | | |
| Cosmetics | High | 352 | -0.05 | 1.02 | 206 | 276 | 0 | $E(p^*_L) > E(p^*_H)$ | *** |
| | Low | 482 | 0.08 | 0.95 | 169 | 183 | 0 | | |
| Shoes | High | 560 | 0.01 | 0.99 | 313 | 247 | 0 | $E(p^*_L) < E(p^*_H)$ | |
| | Low | 385 | -0.01 | 1.01 | 195 | 190 | 0 | | |
| Beverages | High | 526 | 0.10 | 0.94 | 217 | 309 | 0 | $E(p^*_L) < E(p^*_H)$ | *** |
| | Low | 227 | -0.25 | 1.07 | 71 | 156 | 0 | | |

Note: ***, **, * denote that the null hypothesis of zero is rejected at the 1%, 5%, and 10% significance levels respectively.

According to **Table 3**, the relationship varies across different categories. Low-rating sellers charge significantly higher prices compare to high-rating sellers in categories of TVs, laptops, and cosmetics, in agreement with the theoretical hypothesis that low-rating sellers prefer to quit the competition with high-rating sellers and as a result charge a higher price for the profits from uninformed buyers.

The average standardized prices of high- and low-rating sellers are not significantly different, in consistent with the average of standardized price (0.01 and -0.01), and the standard deviation (0.99 and 1.01). This may be due to a larger proportion of bogus products compared to other categories. The presence of fake shoes increases the search cost of potential buyers, as they will need to spend more time on distinguishing genuine brand-name products from fake ones. As a result, buyers are less likely to be well-informed when purchasing these products. The competition for informed buyers is eased, so high-rating sellers charge prices comparable to low-rating sellers. In this case, the negative price premium effect is not observed.

Beverage category is another exception. The average standardized prices of high-rating sellers are significantly greater than low-rating sellers, i.e., a positive price premium effect. The value of beverages is lower than other categories (**Table 1**). Buyers are more likely to arbitrarily make a purchase, instead of comparing offerings to maximize their utility at the cost of time for a low value product.[29] The competition for informed buyers is eased as fewer informed buyers engage the market. Therefore, the negative price premium effect is inversed.

### 4.4 Reputation inflation

As many have noted, feedback ratings on various e-market tend to be implausibly rosy. The median seller on eBay has a score of 100% positive feedback ratings. The tenth percentile is 98.21% positive feedback ratings. On Uber and Lyft, it is widely known that anything less than 5 stars is considered "bad" feedback. Athey et al. stated in Uber, nearly 90% of Chicago trips in early 2017 had a perfect 5-star rating.[30]The main reasons for the online market credit inflation are as follows. Firstly, ratings are prone to inflate as customers may feel pressure to mark sellers down the average, which in turn raise the average score, out of kindness. The standard of buyer satisfaction dropped [30]. Secondly, sellers can manipulate their marks by employing numerous bots or ingenuine buyers who offer positive ratings on fake transactions. [31] Finally, some sellers remove negative comments by deleting them or threatening to retaliate against buyers who give negative review to delete their comment to improve their overall rating. Meanwhile, some enlightening measurements were proposed to control the reputation inflation.

First, sellers could pay for bots or supporters to improve their reputations by rating 'perfect' on fake transactions, which is also called 'score farming'. In view of this, Weijia You et al. proposed a theoretical system to identify characteristics of fake transactions based on the homo economics assumption.[31] They argued transaction-related indicators and customer-related indicators could jointly identify collusive transactions. This system was empirically proved by applying to a database.

The second one is that some sellers remove negative reviews by deleting them or threatening to retaliate against buyers who give negative review to delete their comment to improve the store's reputation. [31] In view of this, the fallback mechanism which was established in the reference 'Use Fallback's online service reputation control mechanism' have some function mainly including four aspects. First, suppress the influence of ingenuine comments or ratings. Second, identify malicious users and punish or quarantine them. Third, motivate buyers to express their real idea. Last, increase the controller's cost (could be money, time, etc.).

The third one is standard of buyer satisfaction dropped. In "reputation inflation", they put forward that since most buyers clearly know the influence of reputation on sellers and worry about whether they will be retaliated if they give reviews in public. Such tolerance and worry will often prompt buyers to give highest score and result in reputation inflation. [30] In view of this, Apostolos et al. proposed an optional anonymous rating system to help reassure buyers so that they can 'speak out' without any worry.

## 5. Conclusions

In this paper, we focus on the pricing strategy of different rated sellers in the online platforms. The pricing strategy influenced by the competition between sellers, different reputation of sellers and the informativeness of buyers. We give the explanation why high reputation sellers may set a lower price comparing with the low reputation sellers in reality, which is contradicted to the literature that the high reputation may seller may enjoy a positive price premium. Here are two reasons for the negative price premium: (1) In the market, there are two types of buyers. The informed buyers tend to know more about the sellers' reputation due to their research on different sellers but they may exist a search cost. The uninformed buyers just randomly choose one seller to buy the product without any search cost. Whether the buyer want to be the informed buyer or uninformed buyer depends on the ratio of search cost and their utility. The co-existing of these two types of buyers may lead to mixed strategies for sellers. What is more, when the proportion of informed buyers exceeds a threshold, it is possible for the high reputation seller to set a lower price than the low reputation seller. (2) There may exist competition between sellers of different reputations. When the competition intensifies, it is impossible for the low reputation seller to get the market of the informed buyers. Hence, they charge the price as high as possible to maximize their profit from the uninformed buyers. Due to the competition among the high reputation sellers, they may reduce their price to compete the market of the informed buyers. Combine these two situations, it is possible that the negative price premium occurs. Our empirical data is also consistent with our theoretical inference of the negative price premium.

We verified the model based on data from Taobao. Sellers adapt different pricing strategy depending on the proportion of informed buyers in the market, which is related to the search cost. Various factors may affect the informativeness of customers. For example, buyers may not invest so much time searching for low value products as for expensive goods; potential fakes may raise the difficulty of searching. The larger proportion of informed buyers in the marketplace, the more intense competitions between high reputation sellers could be. As a result, our findings suggest sellers should adjust their pricing strategies according to their reputation levels and intrinsic attributes of their products.

This research has a number of limitations. Though search cost k is introduced into the model and simulate the expected price at different search costs, little about how to determine the search cost is touched. We assume that low value product and existence of bogus products are mainly responsible for less informative buyers. It would be helpful if a formal approach to quantitively examine the search cost and buyer informativeness across categories could be developed. Moreover, we classify sellers' reputation level as 'high' and 'low' by their median rating. But sellers' ratings are overwhelmingly concentrated around high marks due to reputation inflation. A more delicate classification of sellers may contribute to a more prominent pricing trend. Finally, we define the component rating scores as the reflection of seller's reputation. The binary rating system for each product may also play a role in the pricing strategy but is not considered in this study. Future research may focus on the binary rating system, incorporating text comments, to establish an overall reputation evaluation criterion.

**Appendix A Detailed proof**

**Proof of Proposition 1.** We need three steps to prove Proposition 1:

(1) The domain of setting price for both low and high reputation seller:

Obviously, to keep profit and sales, seller i would set the price $p_i$ between c and $r_i u$. That is c$\leq p_i \leq r_i u$.

More specifically, when c$\leq p_i \leq r_i u$, the $\frac{kn}{\mathrm{u}}$ uninformed buyers go to both sellers for the same chance, and the two sellers compete for market section made up with the $(1-\frac{k}{\mathrm{u}})n$ informed buyers. When seller i offers a higher level of expected utility to buyers ($r_i u - p_i > r_{-i}u - p_{-i}$), he would win all the proportion of informed buyers except for half of uninformed buyers, and gains the profit $(p_i - c)\left[\frac{u-k}{u}n + \frac{(2u-k)n}{2u}\right] = \left(n - \frac{kn}{2u}\right)(p_i - c)$. Otherwise, seller i will only obtain his share of uninformed buyers, which is $\frac{kn}{2\mathrm{u}}$, and gains the profit $\frac{kn}{2u}(p_{-i} - c)$. If $r_i u - p_i = r_{-i}u - p_{-i}$ in some rare case, the two sellers equally get half of total n buyers, and seller i gains the profit $\frac{n}{2}(p_i - c)$.

To win more market share, Seller i is inclined to cut the price $p_i$. However, $p_i$ can be too low that the decrease of united profit outweighs the increase of sales, which means $n - \frac{kn}{2u}(p_i - c) \geq \frac{kn}{2u}(r_i u - c)$. In this case, seller i would give up the market section of the informed buyers and set the price at its upper bound $r_i u$ instead. Thus, we can calculate the lower bound of $p_i$: $P_i \geqslant c + \frac{k}{2u-k}(r_i u - c)$, and we can get a precise domain of $p_i$: $c + \frac{k}{2u-k}(r_i u - c) \leq P_i \leq r_i u$.

Employing this result to both low and high reputation sellers, we can find that seller L's price $P_L$ should be set in the domain $[c + \frac{k}{2u-k}(r_{\mathrm{L}} u - c), r_{\mathrm{L}} u]$, and the counterpart for high reputation seller H should be $[c + \frac{k}{2u-k}(r_{\mathrm{L}} u - c) + (r_{\mathrm{H}} - r_{\mathrm{L}})\mathrm{u}, r_{\mathrm{H}} u]$.

After easy calculation, the minimum profit for seller L is $\frac{k}{2u}n(\widehat{p}_L - c) = \frac{k}{2u}n(r_L u - c)$, and that for seller H is $\lim_{\varepsilon\to 0}\frac{2\mathrm{u}-k}{2u}n(p_H - \varepsilon - c) = \frac{k}{2u} - n(r_H u - c) + \frac{u-k}{u}n(r_H - r_L)u$.

(2) Absence of pure-strategy Nash equilibrium:

First, if $P_H$ is larger than the minimum valid value of $P_L$, the seller L will always try to set $P_L$ lower than $P_H$ with a tiny difference. Thus, there cannot exist a pure-strategy equilibrium.

Second, if $P_H$ is smaller than the minimum valid value of $P_L$, the seller L will have to give up the share of informed buyers and sets $P_L$=$r_L u$. In this case, seller H is willing to increase $P_H$ since he can increase average profit without losing market share.

To conclude, pure-strategy Nash equilibrium could not be reach in this model.

(3) Assuming two Sellers' finally reach the mixed strategies $\{p_L{}^*, p_H{}^*\}$, the expected profit for the high rating seller H at each $p_H$ can be calculate as:

$$\frac{(2u-k)n}{2U}(p_H-c)P(p_L>p_H-(r_H-r_L)u)$$
$$+\frac{n}{2}(p_H-c)P(p_L=p_H-(r_H-r_L)u)$$
$$+-\frac{kn}{2u}(p_H-c)P(p_L<p_H-(r_H-r_L)u)$$

Since the point probability is limited to zero, we can omit the indifference condition and get : $\frac{n}{2}(p_H-c)[1+k-2kP(p_L\le p_H-(r_H-r_L)u)]$

By definition of Nash equilibrium, seller H should be indifference for all his strategies, assuming seller L follows his mixed strategy. Therefore, we can calculate seller H's expected profit by setting his price at the lower-bound price $p_H$ , and get the distribution of $P_L$ . $\frac{n}{2}(p_H^*-c)\left[\frac{2u-k}{u}-2\frac{u-k}{u}F_L^*(p_H^*-(r_H-r_L)\mathrm{u})\right]$

=$\frac{kn}{2u}(r_H u-c)+\frac{u-k}{u}n(r_H-r_L)$

$$\Rightarrow F_L^*(p_L^*)=\frac{(2u-k)p_L^*-k\mathrm{r}_L-2(u-k)c}{2(u-k)(p_L^*+(r_H-r_c)u-c)}$$

$\Rightarrow \overline{F_L}^*$

Here we have $F_L^*(\widehat{p_L})=\frac{r_L u-c}{r_H u-c'}$ which is not equal to 1 as usual. That means the mass probability at upper-bound price for seller L is $1-F_L^*(\widehat{p_L})=\frac{(r_H-r_L)u}{r_r u-c}$. Moreover, we can deduce that seller H will not charge price $\widehat{p_H}$ as high as possible, since the corresponding profit is less than his minimum profit calculated above. The distribution of seller H's setting price can be gain in the same steps.

Done.

**Proof of Proposition 2.** With the same assumptions and similar steps, we can calculate that seller ij's price $p_{ij}$ should be in the range $\left[c+\frac{k}{4u-3\mathrm{k}}(r_i u-c), r_i u\right]$.

Similar to the calculation of pdf in the basic model, the PDF of seller $H_j$ is gained as $\overline{F_{Hj}}^*(p_{Hj}^*)=\frac{k(r_H u-c)}{4(u-k)\left(p_{\mathrm{Hj}}^*-c\right)^2}$, assuming low-reputation sellers set their price at upper bound. This mixed strategy dominates any other strategies of high-reputation sellers. They can not set their price a litter bit lower than $r_L u$ since there are more than one high-reputation sellers just like seller H.

Considering that the high-reputation sellers have large chance to charge a relatively low price due to their optimal mixed strategy, seller $L_j$ would tend to charge the highest possible price $\widehat{p_{Lj}}$ that is,

$$\frac{kn}{4u}(r_L u - c) > \frac{kn}{4u}(p_{Lj} - c) + \frac{u-k}{u} n(p_{Lj} - c)[1 - F(p_{Lj} + (r_H - r_L)u)]^2$$

$$\Leftrightarrow \quad 4\frac{u-k}{u}(p_{Lj} + (r_H - r_L)u - c)^2 + \frac{k}{u}(p_{Lj} - c)(p_{Lj} - r_L u) > 0$$

**Proof of Inference 1.** First, calculating the pricing strategy of both low-and-high-reputation seller and then get their expected price:

$$E(p_L) = \int_{\underline{p_L}}^{\overline{p_L}} p_L \overline{F_L}(p_L) dp_L + r_L u \frac{r_H u - r_L u}{r_H u - c}$$

$$= \frac{(2u-k)(r_H u - c) - 2(u-k)(r_L u - c)}{2(u-k)} \ln\left(\frac{(2u-k)(r_H u - c)}{(2u-k)(r_H u - c) - 2(u-k)(r_L u - c)}\right) + c$$

$$E(p_H) = \int_{\underline{p_H}}^{\overline{p_H}} p_H \overline{F_H}(p_H) dp_H = \ln\left(\frac{2u-k}{k}\right)\frac{k(r_L u - c)}{2(u-k)} + (r_H - r_L)u + c$$

Consider $E(p_H) - E(p_L)$, we have $\frac{d(E(p_H) - E(p_L))}{dk} > 0$, $\lim_{k\to u}(E(p_H^*) - E(p_L^*)) = (r_H - r_L)u > 0$, and $\lim_{k\to 0}(E(p_H^*) - E(p_L^*)) = \left(1 - \ln\left(\frac{r_H u - c}{r_H u - r_L u}\right)\right) \times (r_H - r_L)u$. When $\ln\left(\frac{r_H u - c}{r_H u - r_L u}\right) \geq 1, \lim_{k\to 0}(E(p_H^*) - E(p_L^*)) \leq 0$, there is one and only one turning point $k_1^*$ which satisfies $E(p_H) = E(p_L)$. When $k > k_1^*$, the expected price of the low rating seller will be lower and when $k < k_1^*$, the expected price of the low rating seller will be higher.

**Proof of Inference 2**. In the further competition model, we can calculate the expected price based on pricing strategies in the same way. $E(p_{Lj}) = r_1 u$

$$E(p_{Hj}) = \int p_{Hj} \overline{F_H}(p_{Hj}) dp_{Hj} = c + \frac{k(r_H u - c)}{4(u-k)} \ln\left(\frac{4u - 3k}{k}\right)$$

Then, prove it in similar steps. Consider $E(p_{Hj}) - E(p_{Lj})$, we have $\lim_{k\to u}\left(E(p_{Hj}^*) - E(p_{Lj}^*)\right) > 0, \lim_{k\to 0}\left(E(p_{Hj}^*) - E(p_{Lj}^*)\right) < 0$, and $\frac{d(E(p_{Hj}) - E(p_{Lj}))}{dk} \geqslant 0$. Therefore, there exist one and only one turning point $k_2^*$ satisfying $E(p_{Hj}) = E(p_{Lj})$. When $k > k_2^*$, the expected price of the low rating seller is lower and when $k < k_2^*$, the expected price of the low rating seller is higher.

**Proof of Inference 3**. We need to prove $k_1^* < k_2^*$ to prove inference3.

With the impact of competition among sellers, seller with high reputation reduces the price and seller with low reputaiton choose their highest possible price. we have $E(p_{Hj})(k_1^*) - E(p_{Lj})(k_1^*) > E(p_H)(k_1^*) - E(p_L)(k_1^*)$

If $k_1^* \geq k_2^*$ . Since $\frac{d(E(p_{Hj}) - E(p_{Lj}))}{dk} > 0$ , we can infer $E(p_H)(k_2^*) - E(p_L)(k_2^*) \geq E(p_{Hj})(k_1^*) - E(p_{Lj})(k_1^*)$

Combine the two inequalities, $E(p_H)(k_2^*) - E(p_L)(k_2^*) - E(p_H)(k_1^*) - E(p_L)(k_1^*) = 0$, which occurs a contradiction with $E(p_H)(k_2^*) - E(p_L)(k_2^*) = 0$. Therefore, $k_1^* \geq k_2{}^*$ is impossible.